\documentclass[preprint,showpacs,preprintnumbers,amsmath,amssymb]{revtex4}
\usepackage{amssymb}
\usepackage{amsmath}
\usepackage{graphicx}

\begin{document}

\title{Pseudo-potential Band Structure Calculation of InSb Ultra-thin Films and its application to
assess the n-Metal-Oxide-Semiconductor Transistor Performance}

\author{Zhen Gang Zhu$^{1}$, Tony Low$^{1}$, Ming Fu Li$^{1,2}$,
Wei Jun Fan$^{3}$, P. Bai$^{4}$, D. L. Kwong$^{5}$, G.
Samudra$^{1}$\protect\footnote{eleshanr@nus.edu.sg} }
\address{$^{1}$Silicon Nano Device Lab (SNDL), ECE Department, National University of Singapore (NUS)\\
$^{2}$Fudan University, Shanghai, China 201203 \\
$^{3}$School of EEE, Nanyang Technological University of Singapore\\
$^{4}$Institute of High Performance Computing, Singapore\\
$^{5}$Institute of Microelectronics, Singapore }

\begin{abstract}
Band structure of InSb thin films with $<100>$ surface orientation
is calculated using empirical pseudopotential method (EPM) to
evaluate the performance of nanoscale devices using InSb substrate.
Contrary to the predictions by simple effective mass approximation
methods (EMA), our calculation reveals that $\Gamma$ valley is still
the lowest lying conduction valley. Based on EPM calculations, we
obtained the important electronic structure and transport
parameters, such as effective mass and valley energy minimum, of
InSb thin film as a function of film thickness. Our calculations
reveal that the 'effective mass' of $\Gamma$ valley electrons
increases with the scaling down of the film thickness. We also
provide an assessment of nanoscale InSb thin film devices using
Non-Equilibrium Green's Function under the effective mass framework
in the ballistic regime.
\end{abstract}

 \maketitle

\section{Introduction}
Ultra-thin body (UTB)
metal-oxide-semiconductor-field-effect-transistor (MOSFET) structure
is a promising candidate for scaling MOSFET devices into the
nanometer regime because it has the excellent attribute of
suppressing various short channel effects caused by the downscaling
of device gate length \cite{itrs}. Recently, there are also
experimental and theoretical efforts to evaluate use of
non-conventional channel orientation \cite{yu,komoda,low} or new
channel materials (such as Ge, III-V compound semiconductors, due to
their small $\Gamma$-valley electron masses)
\cite{low,fischetti,ashley} to improve the MOSFET performance.

In particular, InSb is being considered as a new channel material
\cite{ashley} for state-of-the-art MOSFET devices because of its
high bulk mobility (in fact, it has the highest electron and hole
mobility among common III-V semiconductors \cite{mfli,lev}).
Theoretical investigation of InSb MOSFET devices based on a simple
effective mass approximation (EMA) had been conducted in
\cite{pethe} to predict its device performance limits . It was
predicted that the $\Gamma$-valley energy rises rapidly under body
quantization and the inversion charges in the thin film are
transferred to the L-valley. Therefore the advantage of the high
injection velocity from the $\Gamma$-valley electrons is lost. Work
in \cite{pethe} used bulk effective mass even in thin film regime.
However, EMA may not be a reliable method in describing the
electronic band structure of thin films due to unaccounted effects
like band coupling and non-parabolic dispersion. We also found that
the bulk effective mass can only describe the energy dispersion in a
very small range of k space, which brings to question how accurately
EMA can capture the quantization effect in thin film (see the
discussions on Fig. 7 in this paper).

In this work, we describe a more physically reliable and accurate
model, the local empirical pseudopotential method
(EPM)\cite{harrison,heine}, to study InSb thin films. Our results
show that the charge transfer from the $\Gamma$ to the L valley does
not occur and the effective masses in transverse directions become
larger than those in bulk material, which have a direct impact on
the device transport properties. This paper serves to communicate in
detail the models employed in our published conference paper
\cite{zhu}. In section II, we give a detailed description of the
theoretical background of EPM. We will discuss introduction of a
model potential to describe the atomic pseudopotential, which allows
us to extend EPM to thin film calculations. In subsection II. B, we
derive the spin-orbit coupling contribution to the matrix element of
the Hamiltonian used in EPM calculation. In order to achieve
accurate results in EPM calculations of thin film band structures,
we introduce two parameters to take account into the volume
renormalization and the spin-orbit coupling. In subsection II. C, we
discuss the methodology adopted to passivate the surface dangling
bonds by using Hydrogen (H) atom bonding. In section III, we discuss
the important features of InSb thin film electronic structure. We
also obtained important electronic parameters, such as effective
mass and valley energy minimum, of InSb thin film as a function of
film thickness. In section IV, we provide an assessment of nanoscale
InSb thin film double-gated MOSFET devices using non-equilibrium
Green's Function under the effective mass framework in the ballistic
regime derived from realistic thin film band structure reported in
this paper. Finally, we give conclusions in section V.

\section{Theoretical Background}
\begin{figure}[tbh]
\includegraphics[width=6 cm, height=8 cm]{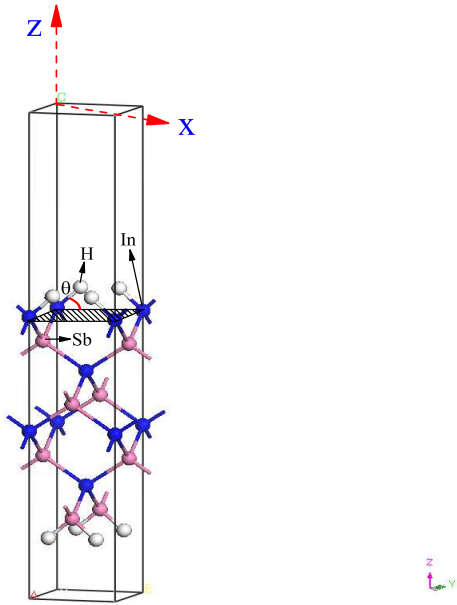}
\caption{(Color online) Supercell of InSb used for our calculation. It
consists of a thin film layer and a vacuum layer in the quantization
direction. $\theta$ is the angle of the H-In bond as illustrated.
The positions of In and Sb atoms are the same as in the bulk
material. $z$ direction is in $<100>$.  \label{fig1}}
\end{figure}
The EPM has already been successfully applied in band structure
calculations of metal, semiconductor or other materials
\cite{fong,cohen,cohen1}. It has also been extended to calculate
band structures and electronic properties of Si quantum dots
\cite{wang,wang1}, quantum wires \cite{xia,yeh}, and quantum films
\cite{xia1,sbzhang}. In view of pedagogical clarity, we shall also
give an overview of the essential theoretical aspects of EPM. Before
going into the rigorous mathematical details, we shall first give an
overall description of the method used in this work. The key aspects
of applying the EPM to a thin film calculation are as follows; (1) a
supercell (Fig. 1) has to be specified prior to any calculation. A
sufficiently large vacuum layer has to be explicitly included in the
supercell so as to eliminate any wavefunction overlap between
different thin film layers. (2) We have to determine the suitable
atomic pseudopotentials for the Sb and In atoms independently for
our thin film calculations. (3) The effect of spin-orbit coupling
needs to be included because it is sufficiently large to affect the
band structure significantly (see Table III) in InSb and many other
compound semiconductors. (4) Hydrogen passivation of the surface
dangling bonds is incorporated into model potential. This is very
crucial treatment as the surface dangling bonds will lead to surface
states in the band gap which will overwhelm the band structure of
the thin film, especially in small band gap materials like InSb.
Therefore we would also need to determine the pseudopotential of
Hydrogen atoms.

\subsection{Pseudopotential Method}
\subsubsection{Empirical Pseudopotential of Bulk InSb}
It is well-known that the band structure can be derived by solving a
secular equation \cite{mfli}. The matrix element of the
pseudopotential (PS) is
\begin{equation}
\langle
PW,\mathbf{G}|\hat{V}^{PS}|PW,\mathbf{G'}\rangle=\sum_{\vec{\tau}}\exp(-i(\mathbf{G}-\mathbf{G}')\cdot\vec{\tau})U_{\vec{\tau}}^{PS}(\mathbf{G}-\mathbf{G}'),\\
\label{vmatrix}
\end{equation}
where
\begin{equation}
U_{\vec{\tau}}^{PS}(\mathbf{G})=\frac{1}{\Omega}\int_{N\Omega}U_{\vec{\tau}}^{PS}(\mathbf{r})\exp(-i\mathbf{G}\cdot\mathbf{r})d\mathbf{r},
\\
\label{ups}
\end{equation}
$\Omega$ is the volume of the unit cell, $N$ is the number of unit
cells, $\mathbf{G}$ and $\mathbf{G}'$ are reciprocal lattice
vectors, $U_{\vec{\tau}}^{PS}(\mathbf{G})$ is the form factor, and
$|PW,\mathbf{G'}\rangle$ is the plane wave basis function. In a
single crystal, there are usually several basis atoms associated
with each lattice point $\vec{r}_{j}$ located at a position in a
unit cell. These atoms have relative positions $\vec{\tau}$ from the
lattice point at $\vec{r}_{j}$. The summation over $\vec{\tau}$ will
yield us the atomic configuration information of the unit cell. Note
that, the bold letters and the letters with an overhead arrow denote
vectors throughout this paper.

\begin{table}[!htp]
\begin{center}
\begin{tabular*}{0.85\textwidth}{@{\extracolsep{\fill}}|c|c|c|c|c|c|c|c|}
\hline \multicolumn{2}{@{\extracolsep{\fill}}|c|} {Form factor (Ry)}
& $q^{2}=0$ & $q^{2}=3$ & $q^{2}=4$ & $q^{2}=8$ & $q^{2}=11$ &
$q^{2}=12$ \\
\hline ESAFF & $U^{S}(q^{2})$ & -0.858 & -0.20 & 0 & 0.018 & 0.034 &
0 \\
& $U^{A}(q^{2})$ & 0 & 0.035 & 0.032 & 0 & 0.011 &
0.013 \\
\hline EMP & $U^{S}(q^{2})$ & -0.816 & -0.202 & 0 & 0.0179 & 0.03416
& 0 \\
& $U^{A}(q^{2})$ & 0 & 0.0353 & 0.0312 & 0 & 0.01221 & 0.0114 \\
\hline
\end{tabular*}
\caption{Symmetry $U^{S}(q^{2})$ and anti-symmetry form factors
$U^{A}(q^{2})$ in our calculations. Unit of $q^{2}$ is
$(2\pi/a)^{2}$, where $a$ is crystal lattice constant. For InSb, we
use $a=6.47877 {\mathrm{\AA}}$. Form factors derived from empirical
model potential (EMP) are also given to establish EMP validity. }
\end{center}
\end{table}

Before we proceed with the calculation of InSb thin film band
structure, we need to derive the atomic pseudopotential (PS) of In
and Sb atoms respectively, or more accurately speaking, to construct
an empirical PS. We can obtain important information about the
atomic PS from bulk InSb band structure, since its calculated band
structure can reliably be calibrated against experimental data. For
bulk InSb, there are two atoms in a unit cell. Conventionally, bulk
empirical PS (at special $q^{2}$ values, they are also known as
Empirical Symmetry and Anti-symmetry Form Factors (ESAFF), as shown
in Table I) can be defined as the summation and the difference of
the individual atomic PSs of Sb and In. The symmetry and
anti-symmetry structure factors are given respectively, as
$S^{S}=\cos(\theta')$ and $S^{A}=\sin(\theta')$, where
$\theta'=\bigtriangleup\mathbf{G}=\mathbf{G}-\mathbf{G}'$. It should
be noted that if we define the symmetry and anti-symmetry form
factors for InSb as $U^{S}=(1/2)(U_{Sb}^{PS}+U_{In}^{PS})$ and
$U^{A}=(1/2)(U_{Sb}^{PS}-U_{In}^{PS})$, then the factor $1/\Omega$
appearing in Eq.(\ref{ups}) will be $1/\Omega_{atomic}$, where
$\Omega_{atomic}$ is the atomic volume in a unit cell.

By performing EPM calculation of the bulk band structure
iteratively, a suitable set of ESAFF is obtained such that it yields
the 'correct' energy gaps at pertinent symmetry points
\cite{mfli,lev}. This set of ESAFF is shown in Table I. Because of
the spherical symmetry approximation of the local pseudopotential,
the first few lowest energy shells are at normalized $q^{2}$ equal
to 3, 4, 8, 11 and 12, where $q$ is wave vector in unit $2\pi/a$ and
$a$ is crystal lattice constant. Based on this set of ESAFF, we can
devise a suitable interpolation scheme that allows us to reasonably
predict a pseudopotential value at other $q^{2}$ values. This will
then allow us to extend this EPM method to calculation of thin film
band structure of arbitrary thickness. An essential assumption we
are invoking is that the atomic pseudopotential is transferable and
thus will not be changed from the bulk to a thin film. This
assumption is reasonable because \textit{ab initio} calculation has
demonstrated that the bulk potential will just be changed in only
one atomic layer at the interface for GaAs/AlAs superlattice
\cite{vande} and will be exactly the same for all other atoms.
Hence, the potential will be bulk-like potential except at one
atomic layer at the surface of the thin film. Therefore, the bulk
atomic potential can be used in thin film calculation. However,
further investigation reveals that we must add the renormalization
factor to the pseudopotential to give the correct results for thin
film as shown in the subsection II. A. 2.

\begin{figure}[tbh]
\includegraphics[width=9 cm, height=4.5 cm]{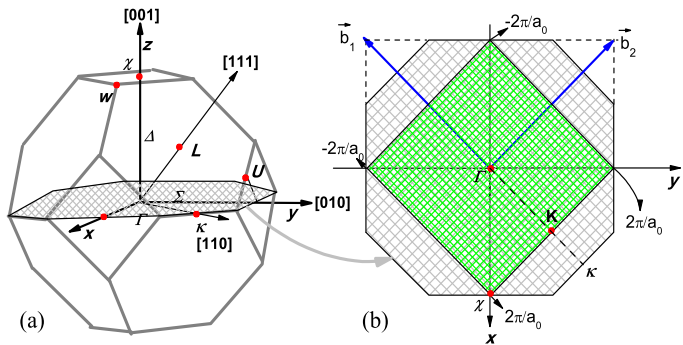}
\caption{(Color online) (a) Three-dimensional Brillouin zone of InSb
and illustration of the important symmetry lines and points
$\Gamma$, $\Delta$, $L$, $X$ and $U$ \cite{mfli}; (b) The
two-dimensional Brillouin zone obtained by projecting onto the xy
plane. The densest shaded region is the first Brillouin zone.  \label{fig2}}
\end{figure}
\subsubsection{Thin Film and Empirical Model Potential (EMP)}
For $<100>$ surface InSb thin film band structure calculation, we
use a supercell as shown in Fig. 1. Fig. (2a) depicts the first
Brillouin zone of bulk InSb and its subsequent projection onto the
xy plane, i.e. the first Brillouin zone of InSb thin film in 2D, is
shown in Fig. (2b). In this paper, band structure of InSb thin film
will be plotted using the same coordinate system for 2D reciprocal
vector space as illustrated in Fig. (2b).

In bulk InSb, band structure is determined by the pseudopotential at
several $q^{2}$ values as stated in previous section and the shells
with $q^{2}<3$ are not of much importance. For thin film InSb, the
reciprocal primitive vector will be shorter than that of bulk and
thus the radius of the lowest energy shell is smaller. This is the
consequence of a larger primitive vector in real space in $z$
direction for $<100>$ film (see Fig. 1). Therefore, the shells with
$q^{2}<3$ are also very important for band structure calculation and
their effects must be properly predicted and included in our
calculation. To account for them, we shall employ an empirical model
potential that can give the continuous pseudopotential as a function
of $q$. We use a model potential \cite{wang,wang1,yeh,sbzhang}
\begin{equation}
V_{atom}^{PS}=\frac{a_{1}(q^{2}-a_{2})}{a_{3}\exp{(a_{4}q^{2})}-1},\\
\label{mp}
\end{equation}
where the subscript "atom" is used to distinguish the different
atomic species' pseudopotential. $a_{1}$, $a_{2}$, $a_{3}$ and
$a_{4}$ are fitting parameters such that they yield the correct form
factor as determined by ESAFF.

\begin{figure}[tbh]
\includegraphics[width=7 cm, height=5 cm]{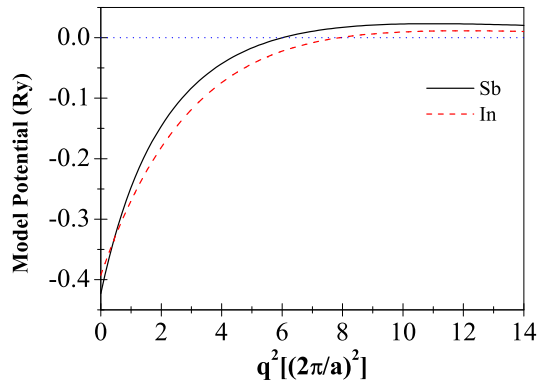}
\caption{(Color online) Atomic model potential of Sb and In. X-axis
and y-axis are in units of $(\frac{2\pi}{a})^{2}$ and Ry (Rydberg),
respectively.  \label{fig3}}
\end{figure}
\begin{table*}[!hbp]
\begin{center}
\begin{ruledtabular}
\begin{tabular}{ccc}
Parameters & Sb & In \\
$a_{1}$ & 0.2588  & 719470  \\
$a_{2}$ & 1.5832  & 2.0811  \\
$a_{3}$ & 1.9689  & 3813600 \\
$a_{4}$ & 0.7159  & 0.9116  \\
\end{tabular}
\end{ruledtabular}
\caption{Derived parameters for model potentials of atoms Sb and
In.}
\end{center}
\end{table*}

To obtain a unique atomic model potential of In and Sb, we need
eight constraints for the eight fitting parameters. Seven
constraints come from the seven nonzero ESAFF parameters at $q^{2}$
equal to 3, 4, 8, 11 and 12 shown in Table I. The remaining one
constraint is imposed by InSb work function (WF), which gives one
nonzero ESAFF at $q^{2}$ equal to 0 \cite{sbzhang}. This last
constraint accounts for the surface property, which is important for
a thin film \cite{sbzhang}. In fitting these parameters, it should
be noted that the physical unit of $q^{2}$ in Eq. (\ref{mp}) is
$(2\pi/a)^{2}$ and $V_{atom}^{PS}$ is in rydberg atomic unit. The
best fitting parameters for the model potential are shown in Table
II and its variation with $q^{2}$ is shown in Fig. 3. From this
figure, we observe that the model potentials of Sb and In atoms are
in a reasonable range. Also, the fits to $U^{S}$ and $U^{A}$ are
quite good as is clear from Table I.

\begin{table*}[!hbp]
\begin{center}
 \vspace{0.7cm}
\begin{ruledtabular}
\begin{tabular}{c|c|c|c}
Energy (eV) & Experiment & Calculation 1 (ESAFF) & Calculation 2 (EMP)\\
\hline $E_{\Gamma}$ & 0.17 & 0.172 & 0.168 \\
\hline $E_{L}$      & 0.68 & 0.685 & 0.704 \\
\hline $E_{X}$      & 1.0  & 0.995 & 1.034 \\
\hline $E_{SO}$     & 0.8  & 0.801 & 0.802 \\
\hline WF           & -4.76 & -4.760 & -4.223\\
\end{tabular}
\end{ruledtabular}
\caption{Experimental data \cite{mfli,lev} of various energy
conduction valley minima ($E_{\Gamma}$, $E_{L}$ and $E_{X}$),
spin-orbit coupling ($E_{SO}$) and work function (WF) are compared
with theoretical result from empirical pseudopotential method.
Calculation 1 and 2 employs the Empirical Symmetry and Anti-symmetry
Form Factors (ESAFF) and Empirical Model Potential (EMP)
respectively shown in Table I. Although the error of WF derived from
EMP is large, the relative error is still tolerable.}
\end{center}
\end{table*}

We perform EPM calculation of bulk InSb using this new model
potential and reproduce the required energy band minima derived from
experiment, as tabulated in Table III. We shall point out that the
model potential of one of the species of an atom in a semiconductor
compound is actually atomic model potential, i.e. the volume of unit
cell is divided by the number of atoms in this unit cell. So for the
thin film calculations, there will be a renormalization with a
different unit cell volume under the assumption that the overall
pseudopotential of atoms is the superposition of the local
pseudopotential of all the atoms in the thin film.

\begin{figure}[tbh]
\includegraphics[width=8 cm, height=6 cm]{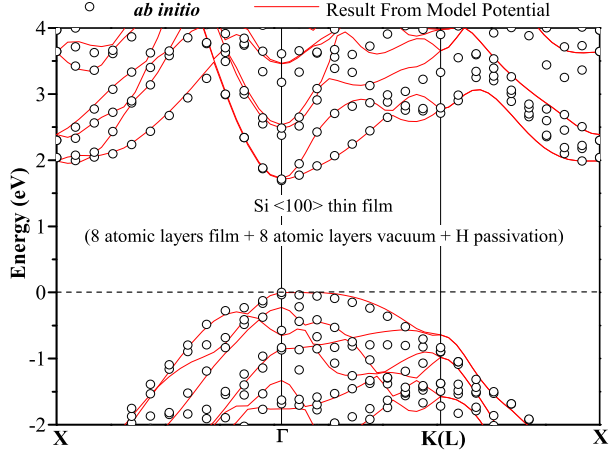}
\caption{(Color online) Energy bands of Si $<100>$ thin film (8 atomic
layers Si film + 8 atomic layers vacuum with H passivation)
calculated with ab initio method (using CASTEP, Local Density
Approximation) and via empirical pseudopotential method with a model
potential. Excellent agreement for the various conduction valleys
minima is obtained.  \label{fig4}}
\end{figure}
\subsection{Hydrogen Passivation}
In thin film calculation, we must ensure that surface dangling bonds
are correctly passivated with Hydrogen (H) so that there will be no
surface states within the band gap. To determine model potential for
Hydrogen, we tried model potential of H in \cite{wang,yeh,sbzhang}
and calculated the band structure of 8 atomic layers (atm) Si
$<100>$ thin film and compared the EPM results with those derived
from \textit{ab initio} calculation \cite{segall} (Fig. 4). The band
structure obtained from both methods are very consistent, thus
establishing the validity and reliability of this H model potential.
In Si $<100>$ thin film calculation, the renormalization of
supercell volume is set as a ratio of the number of H atoms in the
supercell and the number of Si atoms in the same unit cell. Also we
assumed that the atomic volume of H is the same as that of Si.


\begin{figure}
\centering
\includegraphics[width=5 cm,height=4.7 cm]{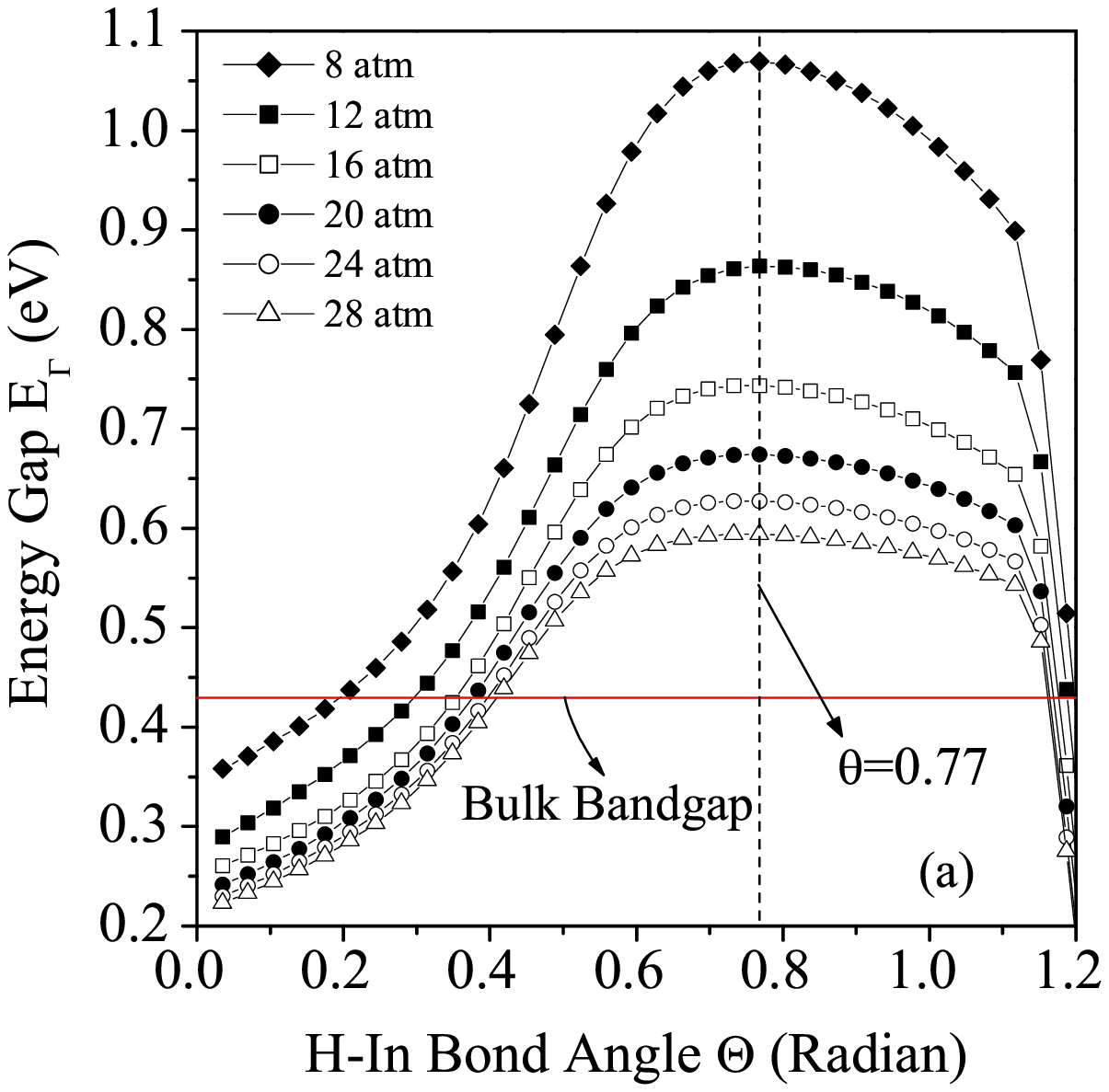}
\includegraphics[width=5 cm,height=5.2 cm]{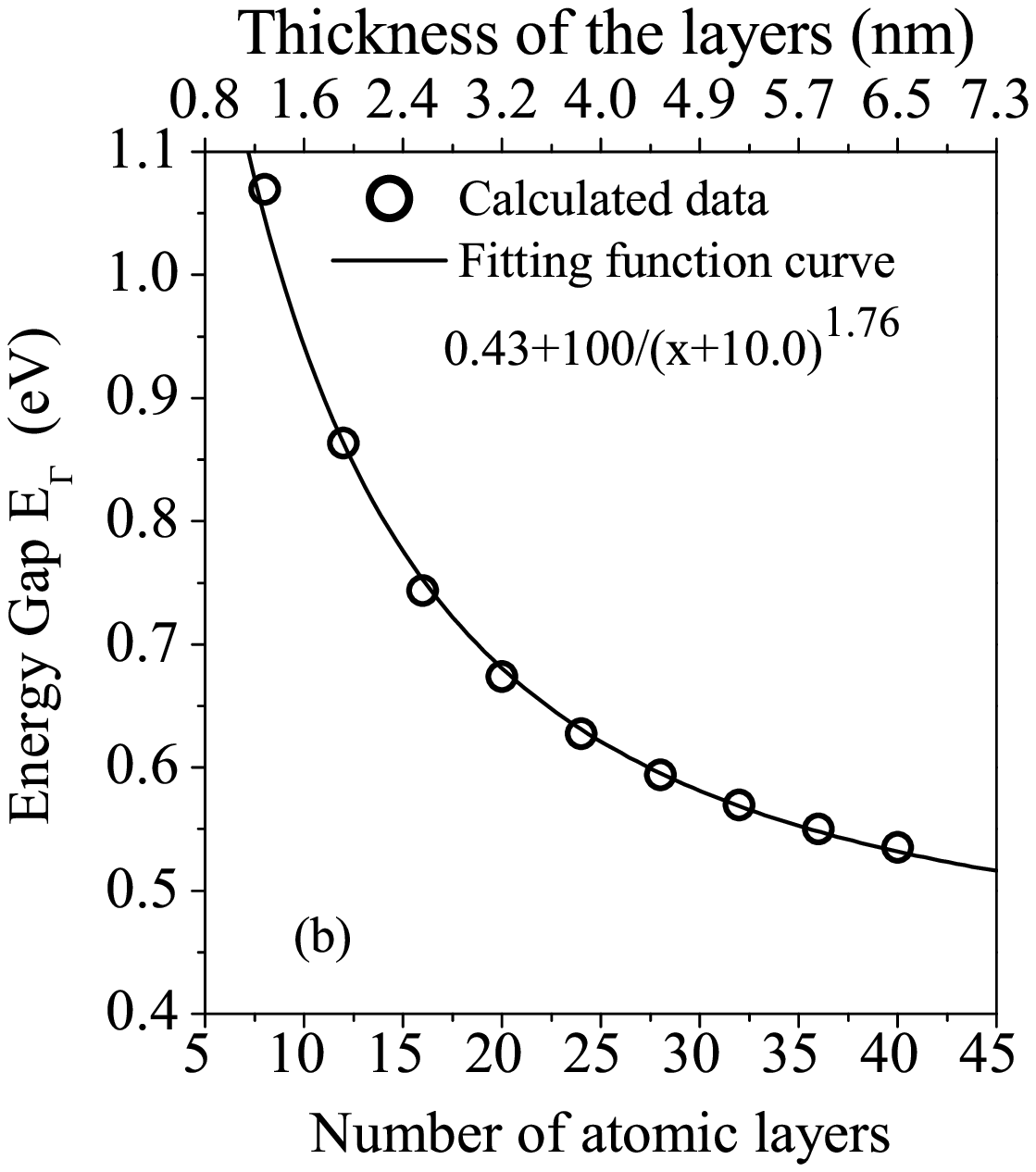}
\caption{(Color online) (a) The calculated InSb direct energy gap
($E_{\Gamma}$) as a function of H-In bond angle $\theta$, plotted
for InSb thin films of varying number of atomic layers (atm). The
graph illustrates that there is a common $\theta$ where the energy
function is a stationary point. (b) The direct energy gap
($E_{\Gamma}$) at bond angle $\theta$ where the energy function is a
stationary point, plotted as a function of film thickness. An
expression which fits these data points is given. It gives correct
asymptotic behavior when film thickness tends to bulk, yielding bulk
band gap value of 0.43 eV.}
\label{fig5}
\end{figure}

For InSb thin film, the bond length of In-H is set as $1.75
\mathrm{\AA}$ (from experiment in \cite{rag}) and Sb-H as $1.711
\mathrm{\AA}$ (from \textit{ab initio} energy minimization
calculation \cite{segall}, which is very close to $1.67
\mathrm{\AA}$ in Ref. \cite{gabor}). The angle between the bond of
In-H and the top surface ($\theta$) (Fig. 1), should be determined
carefully. Fig. (5a) shows the calculated direct InSb band gap
($E_{\Gamma}$) as a function of $\theta$ for varying number of InSb
atomic layers. In principle, the energetically stable $\theta$ for
the thin film system is when $E_{\Gamma}$ is stationary with respect
to $\theta$. $E_{\Gamma}$ as a function of $\theta$ shown in Fig.
(5a) confirms $\theta=0.77 \text{rad}$ as a stationary value for
$\theta$ and a fitting formula (similar to \cite{wang,yeh,sbzhang})
affirms that $E_{\Gamma}$ approaches the bulk band gap of $0.43$ eV
as the film thickness tends to the bulk limit as shown in Fig. (5b)
(without spin-orbit contribution).

For InSb thin film, we need to renormalize Hydrogen atomic potential
derived from Si thin film calculation,
$V_{H}^{PS}(\text{InSb})=(\Omega_{\text{Si}}/\Omega_{\text{InSb}})V_{H}^{PS}(\text{Si})$
using relative volume of H atoms compared to the volume of InSb,
where $\Omega_{\text{Si}}$ and $\Omega_{\text{InSb}}$ are supercell
volumes of Si and InSb respectively. Since the InSb thin film has
the same supercell structure as Si thin film, its supercell
renomalization is performed using their respective lattice constants
as
$(\Omega_{\text{Si}}/\Omega_{\text{InSb}})=(a_{\text{Si}}/a_{\text{InSb}})^{3}$.

\subsection{Spin-Orbit Coupling (SOC)}
SOC treatment is incorporated by calculating the matrix element of
the spin-orbit coupling term in the Hamiltonian expressed in a plane
wave representation. It can be reduced to an expression similar as
the pseudopotential form factor, known as the spin orbit form
factors (SOFFs) \cite{weisz,che,walter,bloom}. In this section, we
will clarify the formulation used for SOC treatment. A more
mathematically detailed derivation shall be presented in the
Appendix.

If we consider a supercell composed of different species of basis
atoms at each lattice site, the spin-orbit Hamiltonian can be
specified as
\begin{equation}
H_{\vec{k'}s',\vec{k}s}^{SO}=-i\vec{\sigma}_{s's}\cdot(\vec{k'}\times\vec{k})\sum_{j=1}^{m}S_{j}(\vartriangle\vec{k})\lambda_{j},\\
\label{hsoso3}
\end{equation}
where $\lambda_{j}$ is a parameter for $j$-th species, $m$ is the
number of species, $\vartriangle\vec{k}=\vec{k'}-\vec{k}$ is the
change in the wave vector, $\vec{\sigma}_{s's}$ are matrix elements
of Pauli matrices, $s$ and $s'$ denote the spin states; up spin or
down spin. Structure factor is given as
$S_{j}(\vartriangle\vec{k})=n_{num}^{-1}\sum_{n_{j}}\exp(i\vartriangle\vec{k}\cdot\vec{\tau}_{n_{j}})$,
where the summation is over $n_{num}$ atoms of $j$-th species (see
Appendix for more detail).

\begin{figure}[tbh]
\includegraphics[width=8 cm, height=4.6 cm]{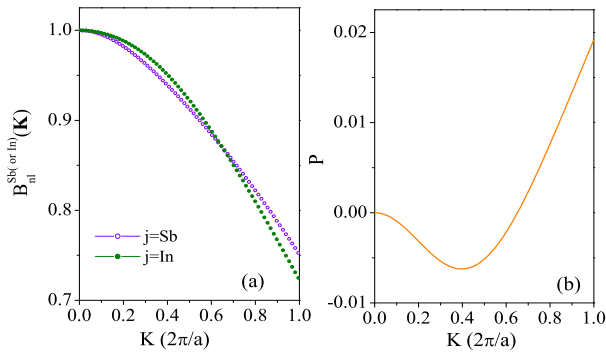}
\caption{(Color online) (a) Variation of $B$ function with wave vector
for Sb atom and In atom for $n=4$ and $l=1$. (b) Its relative
difference $P$ with wave vector, where
P=($B^{Sb}_{nl}$-$B^{In}_{nl}$)/($B^{Sb}_{nl}$+$B^{In}_{nl}$).  \label{fig6}}
\end{figure}

The parameters $\lambda_{j}$ are determined by the $B$ function
defined in the appendix (see Eq. (\ref{rdb})). The spin-orbit
parameter $\eta$ (see Eq. (\ref{eta})) is a property of isolated
atoms, so it is independent of the crystal wave vector. The defined
$B$ function is important in the calculation of $\lambda_{j}$ and
its behavior is depicted in Fig. 6.

In Fig. 6, we plot the variation of the B function (i.e. Eq.
(\ref{rdb})) as a function of wave vector for Sb and In atoms in (a)
and the difference of the B function between these two types of
atoms in (b). For our plot in Fig. 6, we just show the outermost
core state, i.e. $n=4$ and $l=1$ ($p$ orbit). Because of the
normalization, the B function tends to 1 as $k\rightarrow 0$.

Although the explicit form of the spin-orbit coupling matrix element
is presented in the Appendix, there are some additional steps that
must be taken into consideration in order to obtain the correct
result. Thus, we shall briefly elaborate the procedures involved in
SOC treatment. First off all, the SOFFs are obtained via a similar
methodology as ESAFF; by iterating our EPM calculations until the
correct energy gaps are obtained for bulk InSb. SOFFs have to be
normalized with respect to supercell volume when used for a thin
film calculation. Besides the usual volume renormalization, one has
to account for another renormalization. Conventionally, bulk band
structure of InSb is derived by considering a 2-atom unit cell in
which the primitive vectors are non-orthogonal and the mutual angles
between them are all $\pi/3$. In thin film calculation, a larger
supercell is constructed in which the primitive vectors are mutually
orthogonal. Hence this angular effect renormalization and volume
renormalization must be considered in our treatment.

\begin{figure}
\centering
\includegraphics[width=5 cm,height=5 cm]{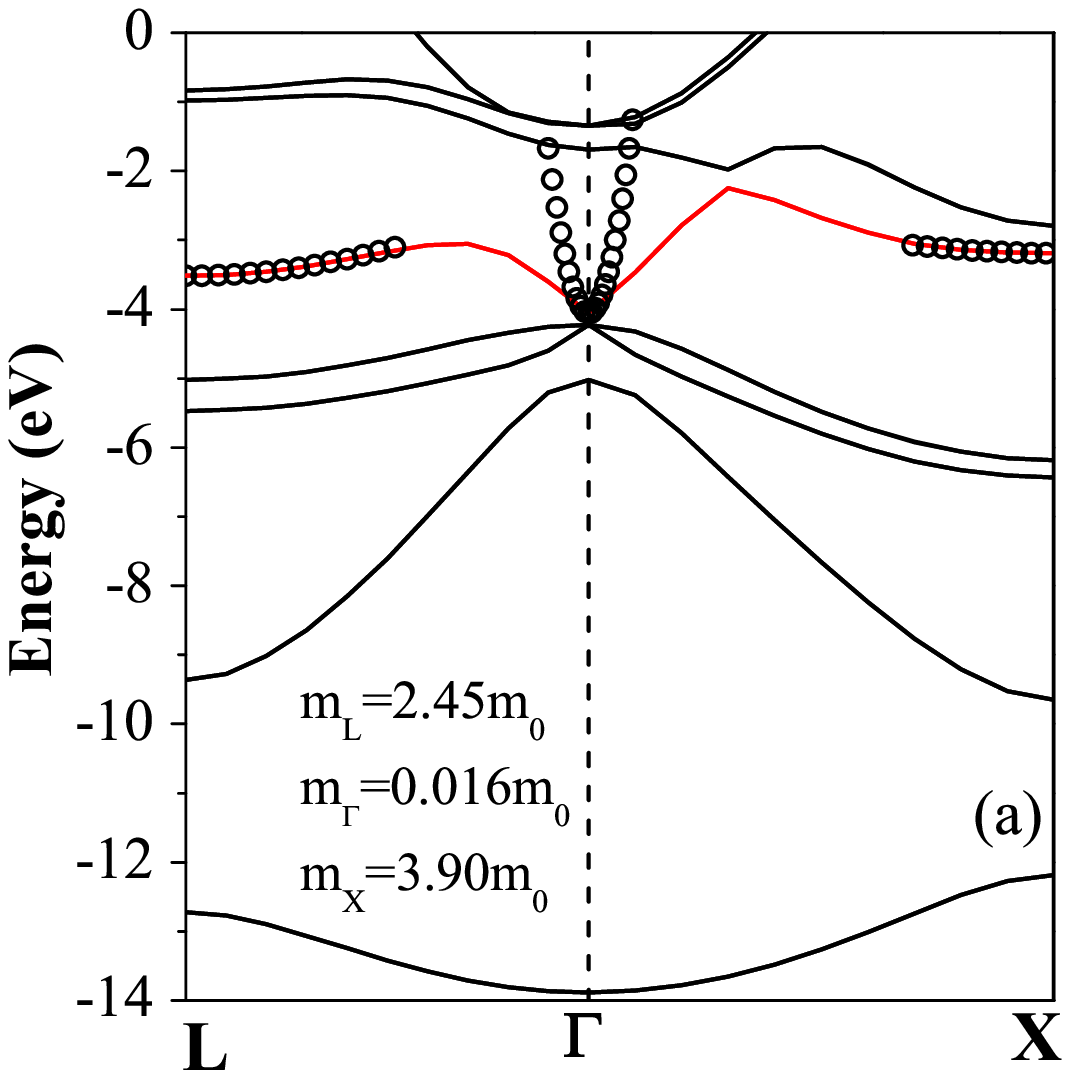}
\includegraphics[width=5 cm,height=5 cm]{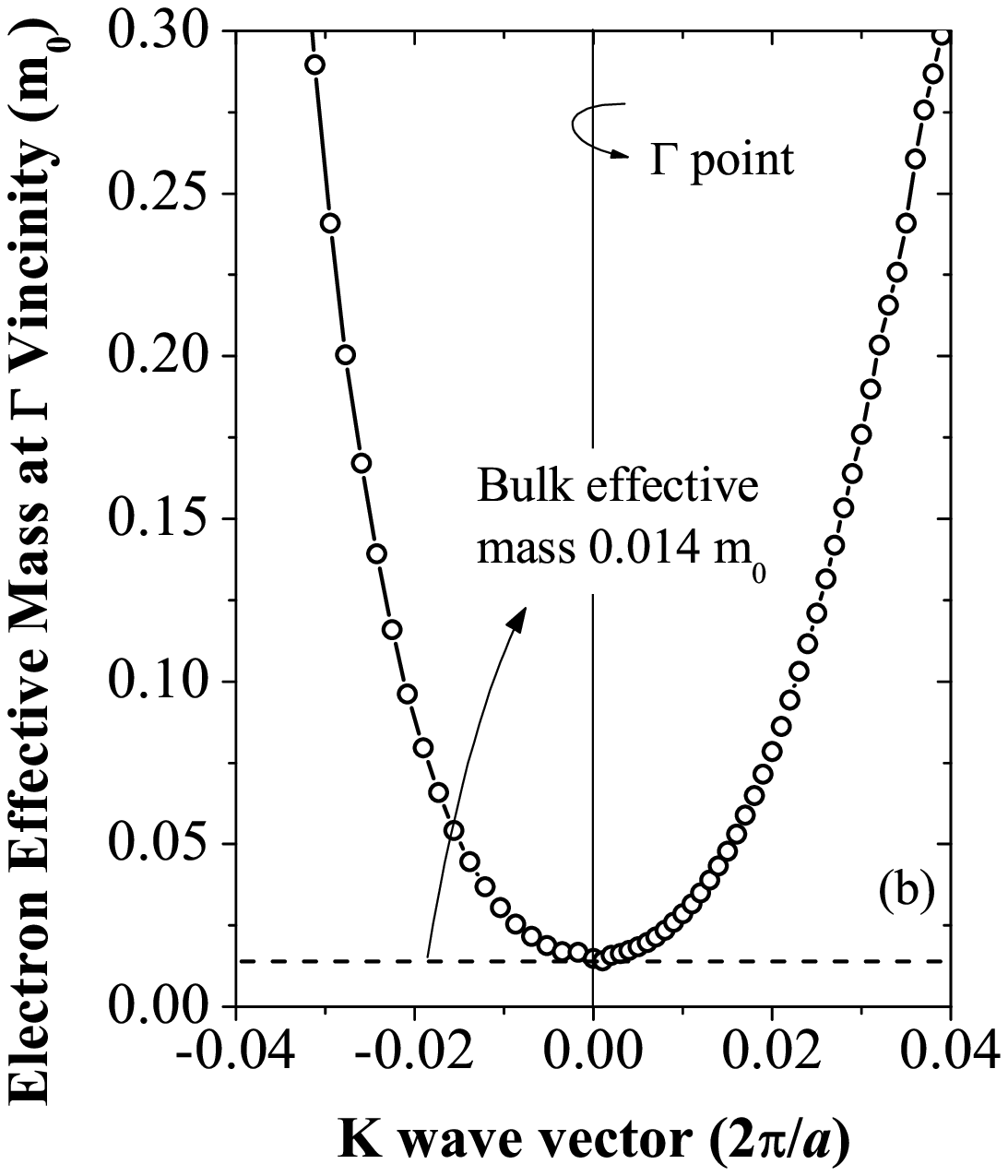}
\caption{(Color online) (a) Energy band structure of bulk InSb
calculated using empirical pseudopotential method. We have
calculated the various effective masses for the different conduction
valleys. The longitudinal mass (derived from the empty circles in
(a) at L, X and $\Gamma$) for L and X valleys are found to be
$m_{L}=2.45m_{0}$ and $m_{X}=3.90m_{0}$, respectively. Lastly, the
isotropic effective mass at $\Gamma$ valley is found to be
$m_{\Gamma}=0.016m_{0}$. (b) Effective mass of $\Gamma$ valley
($m_{\Gamma}$) calculated from the second derivative of energy
dispersion.}
\label{fig7}
\end{figure}

The spin orbit Hamiltonian is divided into two parts; one part
dealing with the angular part and the other the magnitude part. When
the supercell is changed from bulk 2-atom unit cell to $<100>$ thin
film supercell, the volume renormalization will appear in the second
part like what has been done on form factors of pseudopotential of
Sb and In. For the angular part renormalization, we treat it as an
unknown parameter which we derived using the following procedure. We
removed the vacuum slab from the thin film supercell. This new
supercell is therefore a unit cell for a bulk crystal. We then
adjust the angular part renormalization parameter so that it
reproduces the correct bulk band structure and spin orbit splitting
(as calculated from a conventional bulk crystal unit cell). This
procedure has to be repeated for each film thickness. After this
remormalization, the vacuum slabs are replaced again and the thin
film band structure can be derived from EPM.

\begin{figure}[tbh]
\includegraphics[width=10 cm, height=7 cm]{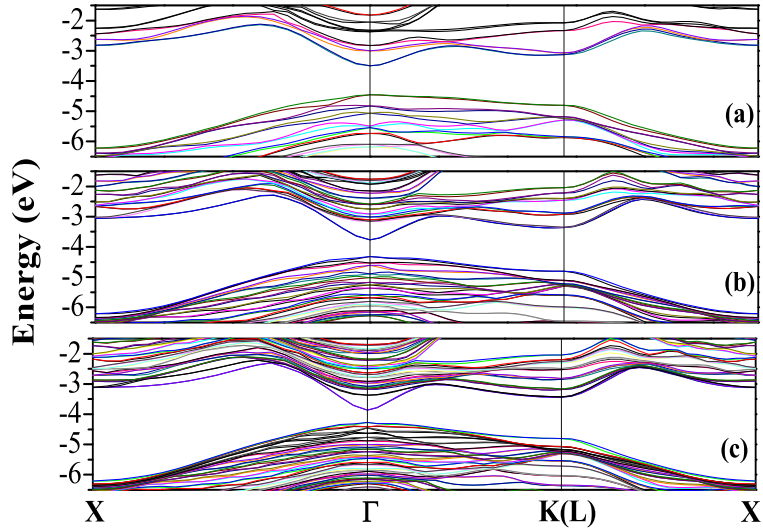}
\caption{(Color online) Band structure of InSb thin film. There are 8
atm layer, 16 atm layer and 24 atm layer film in (a), (b) and (c),
respectively.  \label{fig8}}
\end{figure}

\section{Features of I\lowercase{n}S\lowercase{b} Electronic Structures}
Fig. (7a) shows the calculated band structure of bulk InSb from our
EPM. We derived the following effective masses: $m_{L}=2.45m_{0}$
(the longitudinal mass of L valley), $m_{X}=3.90m_{0}$ (the
longitudinal mass of $\Delta$ valley) and $m_{\Gamma}=0.016m_{0}$
(the isotropic mass of $\Gamma$ valley). The open circles in Fig
(7a) are data points fitted to the bands minima by using a parabolic
dispersion with effective masses as stated. Fig. (7b) shows the
'effective mass' in the vicinity of $\Gamma$ valley ($m_{\Gamma}$)
calculated by taking the second derivative of energy with respect to
the wave vector $k$. We note that the parabolic assumption for the
energy dispersion at $\Gamma$ is only valid for a very small $k$
range. Consider a 1.3 nm InSb film, the wave vector spread according
to uncertainty principle is $\sim0.04(2\pi/a)$. From the plot, the
parabolic assumption for EMA is only applicable up to a $k$ range of
$\sim0.01(2\pi/a)$. Hence, in the ultra thin film regime, a
parabolic EMA is not a reasonable assumption and becomes a highly
unreliable method for calculation of size quantization effect. In
addition, the small band gap also entails a considerable amount of
coupling between the conduction and valence bands, which render the
uncoupled EMA approach unreliable.

\begin{figure}
\centering
\includegraphics[width=5 cm,height=5 cm]{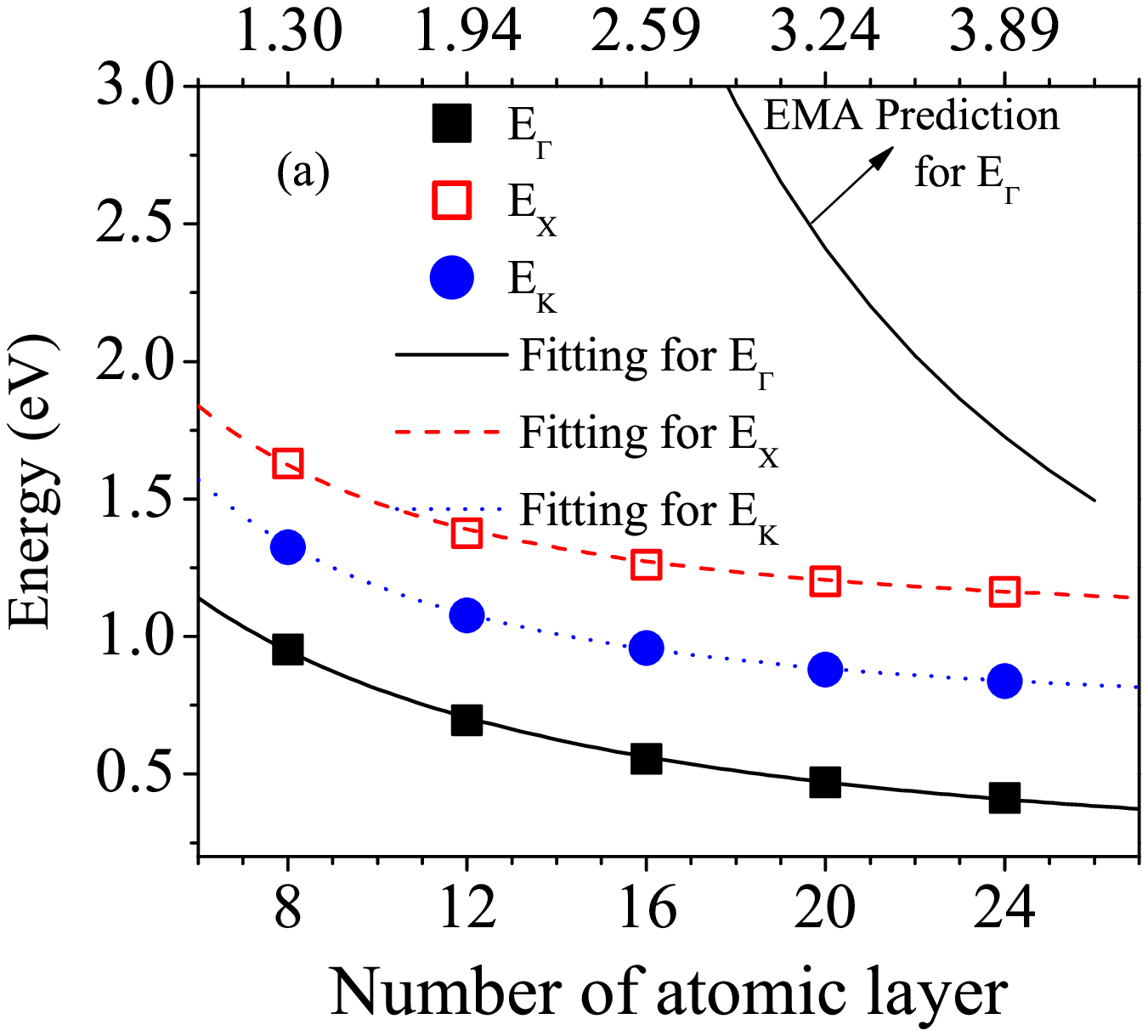}
\includegraphics[width=5 cm,height=5 cm]{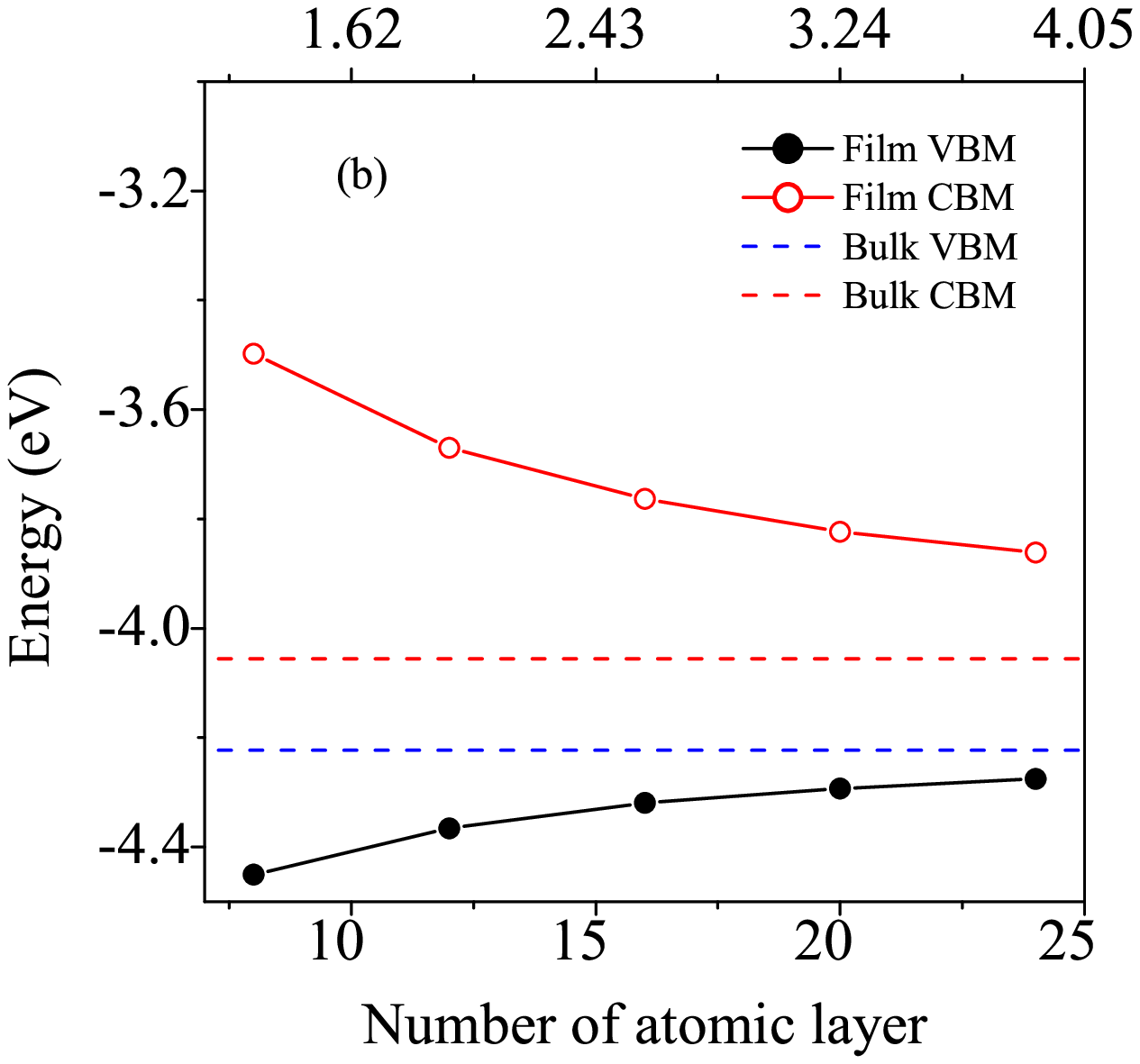}
\caption{(Color online) (a) Variation of band gap with the number of
atomic layers. (Top scale shows the thickness of thin film in nm in
(a) and (b)). (b) The InSb thin film valence band maximum (VBM) and
conduction band minimum (CBM) as a function of the number of atomic
layers. Bulk values are approached at thicker films.}
\label{fig9}
\end{figure}

Fig. 8 shows the calculated band structure of thin film InSb from
our EPM. These thin films still retain the direct band gap
properties in contrast to EMA calculation \cite{pethe}, which has
$L$ valley as the lowest lying conduction valley when film thickness
is below 5 nm. Fig. (9a) shows variation of the energy minima at
$\Gamma, L$ and $\Delta$ valleys as function of film thickness. We
also plot the fitting curves for the energy minima as a function of
thickness x in this figure. The fitting formulas are
$0.16802+150/(x+9.7)^{1.83}$, $1.03395+92/(x+5.5)^{1.94}$, and
$0.70439+94/(x+5.2)^{1.94}$, respectively. There was no crossing of
the energy minima down to 6 atomic layers, in contrast to
predictions by EMA methods. In addition, we note that the lowest
lying $\Gamma$ valley is separated from the other valleys with a gap
more than $0.3$ eV, signifying that electrons are dominantly
occupying $\Gamma$ valley. Another starking contrast with results of
EMA is the band gap. Although, the band gap is enhanced by
quantization effect, this increase with reducing thickness is much
slower than the well-known $d^{-2}$ behavior predicted by EMA in an
infinite deep well model (For e.g. at $\Gamma$ valley, see Fig.
(9a)). However, EPM also yields the result that the conduction band
quantization effect is larger than that of valence band in agreement
with prediction by EMA (Fig. (9b)). From a logic device
point-of-view, one would desire to have a larger band gap to curb
the increasing band-to-band tunneling current with each new
generation CMOS devices. The fact that InSb is a direct band gap
material may aggravate the problem. However, if one can achieve a
sufficiently thin film with $\sim8$atm, which offers a band gap more
than $1$ eV, band-to-band tunneling current should still be
tolerable.

\begin{figure}[tbh]
\includegraphics[width=10 cm, height=5 cm]{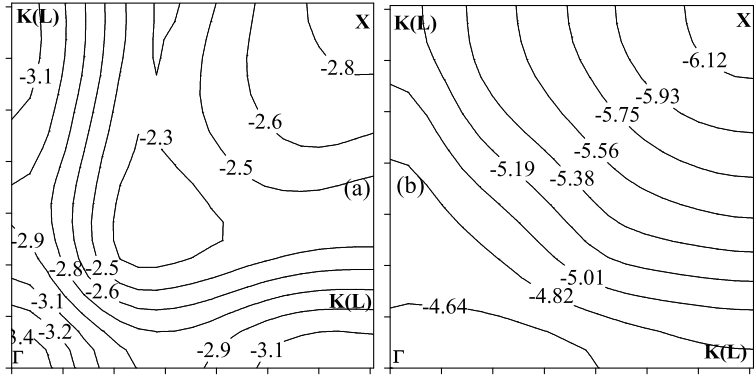}
\caption{Energy contour plot for 8 atomic layers InSb thin film for
the first conduction and first valence band are plotted in (a) and
(b), respectively. Values of the energy are indicated on each
contour line. Important symmetry points are also indicated. It is
apparent that the $\Gamma$ conduction valley is relatively isotropic
in nature even in the ultra-thin film regime.  \label{fig10}}
\end{figure}

Fig. 10 shows the contour plot of the conduction and valence band
energy dispersion for 8 atomic layers InSb thin film. We confirm
that the $\Gamma$ valley in thin film still retains its isotropy
unlike the case of Si \cite{low}. Fig. 11 shows the Si and InSb thin
film effective mass fitted from their 2D energy dispersion. InSb
$\Gamma$ valley is isotropic with increasing effective mass as film
thickness is scaled down. Si $\Gamma$ valley in thin film actually
is originated from bulk $\Delta$ valley, projected onto the 2D
k-space. We observe that anisotropy in Si becomes more prominent
with decreasing of film thickness, with the $\Gamma K$ direction
effective mass diverging for each of the two degenerate bands. The
larger $\Gamma K$ effective mass in Si is for the lower band of the
two degenerate bands, and the smaller $\Gamma K$ effective mass
corresponds to the top band.

We have attempted to fit a parabolic dispersion to match the
calculated energy dispersion to obtain the effective mass. In fact,
we expect the effective mass to also increase with wave-vector as
depicted in Fig. (7b). However, even if effective mass is not a
rigorously derived quantity due to non-parabolic valleys, it serves
as a very useful 'figure of merit'. Interestingly, the isotropic
effective mass of InSb $\Gamma$ valley increases with the decrease
of film thickness. At $\sim1$ nm of InSb film, the effective mass is
$\sim0.1m_{0}$. Increase in $m_{\Gamma}$ would retard the
quantization effect in InSb thin film. The isotropy of the electron
mass, which translates to isotropy of the electron transport
property, should be advantageous as it affords engineer with more
flexibility in orientating the n and p-MOSFET devices to yield the
most optimum transport direction on the same substrate.

\begin{figure}[tbh]
\includegraphics[width=8 cm, height=6.3 cm]{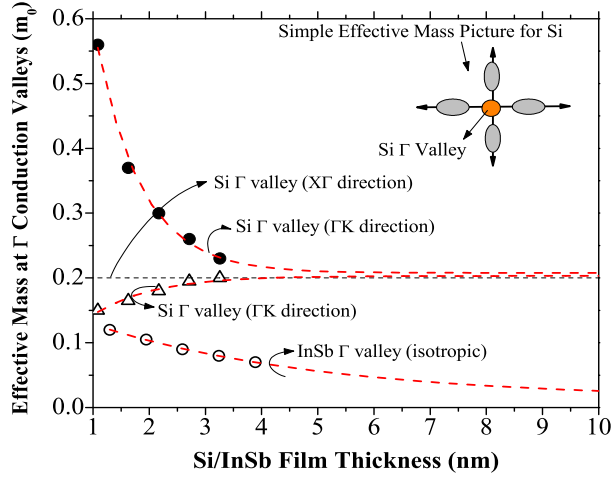}
\caption{(Color online) Electron effective mass of Si and InSb
$\Gamma$ conduction valley. The inset shows the constant energy
ellipsoid for Si projected onto the 2D k-space, with the $\Gamma$
valley being doubly degenerate.  \label{fig11}}
\end{figure}
\begin{figure}[tbh]
\includegraphics[width=8 cm, height=6.3 cm]{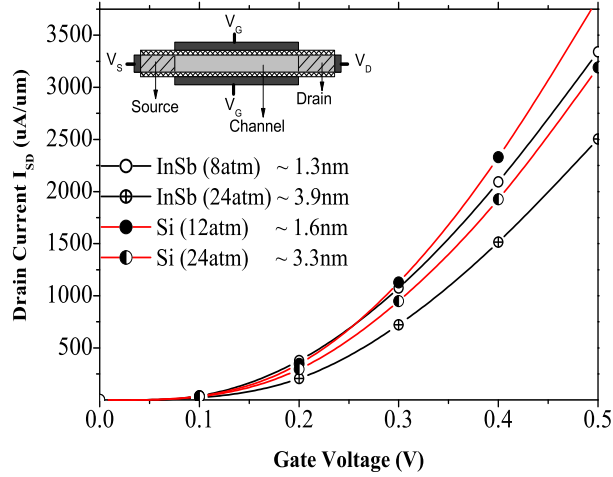}
\caption{(Color online) Ballistic current calculated for InSb and Si
double-gated devices using simple effective mass approximation. 8atm
and 24atm InSb/Si thin film devices are considered. Effective masses
used for InSb are derived from a parabolic fit of the band structure
from EPM.  \label{fig12}}
\end{figure}

\section{Ballistic Limit of I\lowercase{n}S\lowercase{b} Nanoscale Devices}
We calculated the InSb thin film devices' ballistic current limit
using the Non-Equilibrium Green's Function (NEGF) method \cite{ren},
under the framework of simple effective mass method. Since the
energy dispersion for $<100>$-surface InSb thin film is relatively
isotropic (Fig. 10), one can employ a decoupled 2D treatment to the
problem and calculation is done in the framework of mode-space NEGF
approach \cite{ren}. The effective masses used for the various InSb
devices are as derived in Fig. 11. For Si devices, due to the
anisotropy, we simply assumed an effective mass of $0.20m_{0}$. A
double-gated device structure shown in Fig. 12 is employed. Channel
length of 20 nm is used. An EOT of 1 nm and metal gate is employed.
No wave function penetration into oxide is assumed. N-type
Source/drain and p-type channel are doped at $1\times10^{20}cm^{-3}$
and $1\times10^{15}cm^{-3}$ respectively. The boundary conditions
for the potential at the contact are set to floating, by imposing
the condition that the potential derivatives are zero at these
boundaries. The drive current vs. gate voltage is plotted in Fig.
12. The Si devices yield a slightly larger ballistic current
compared to their InSb counterparts. The main reason is due to the
larger density-of-states mass in Si and the fact that it is doubly
degenerate compared to non-degenerate InSb $\Gamma$ valley.
Currently, experimental study shows that sub 50 nm Si devices only
achieve $\sim40\%$ ballistic performance \cite{loc}. A lower
density-of-states mass should help damp the dissipative processes to
achieve fully ballistic performances in InSb devices and performance
then could be better than Si.

\section{Conclusion}
Band structure of III-V material InSb thin films is calculated using
empirical pseudopotential method (EPM). $\Gamma$ valley in InSb
remains the lowest lying conduction valley (a desirable trait if
high mobility characteristic is required) despite size quantization
effects but its isotropic effective mass increases with decrease in
the film thickness. A NEGF calculation of the InSb and Si
double-gated devices reveals that they have comparable ballistic
drive current.

\section*{Acknowledgments}
This work is supported by Singapore A*STAR Nanoelectronics Program
under research grants R-398000019305. We gratefully acknowledge the
use of NanoMOS \cite{ren} from Purdue
Computational Electronics Group for this work.\\

\appendix

\section{treatment of spin-orbit coupling}

The first starting point is the formulas in Ref. \cite{weisz}, such
as equation (A3) in it. To explicitly write down the formula of SOC,
we shall calculate one of the terms of Eq. (A4) in Ref. \cite{weisz}
denoted as (II) here. We consider
\begin{eqnarray}
(II)&=& -\sum_{t}\langle\vec{k'}|b_{t}\rangle\langle
b_{t}|\vec{\Lambda}|\vec{k}\rangle \notag\\
&=&-\frac{1}{N_{cell}\Omega_{cell}}\sum_{t=nlm}\sum_{ij}\int
d\vec{r}_{1}d\vec{r}_{2}d\vec{r}_{3}e^{-i\vec{k'}\cdot\vec{r}_{1}}\psi_{nlm}(|\vec{r}_{1}-\vec{R}_{j}-\vec{\tau}_{i}|,\theta_{1},\phi_{1}) \notag\\
&&\times\psi_{nlm}^{*}(|\vec{r}_{2}-\vec{R}_{j}-\vec{\tau}_{i}|,\theta_{2},\phi_{2})\xi(|\vec{r}_{3}-\vec{R}_{j}-\vec{\tau}_{i}|)
\vec{l}(\vec{r}_{3}-\vec{R}_{j}-\vec{\tau}_{i})\delta(\vec{r}_{2}-\vec{r}_{3})e^{i\vec{k}\cdot\vec{r}_{3}}
\label{ii}
\end{eqnarray}%
where $n,l$ and $m$ are quantum numbers characterizing the core
states, $t=nlm$, $\vec{R}_{j}$ is the point located the unit cell,
$\vec{\tau}_{i}$ are the coordinates of atoms in the basis,
$\xi(|\vec{r}|)=\frac{1}{4mc^{2}}\frac{1}{|\vec{r}|}\frac{dV(|\vec{r}|)}{d|\vec{r}|}$,
with $V$ being potential of atomic nuclei, $\vec{l}$ is the angular
momentum operator, $\delta$ is the Dirac $\delta$ function,
$\Omega_{cell}$ is the volume of the unit cell, and $N_{cell}$ is
the number of unit cells in the crystal. By performing the integral
over $\vec{r}_{3}$, and then let
$\vec{r}_{1}^{\hspace{0.1cm}\prime}$=$\vec{r}_{1}-\vec{R}_{j}-\vec{\tau}_{i}$
and
$\vec{r}_{2}^{\hspace{0.1cm}\prime}$=$\vec{r}_{2}-\vec{R}_{j}-\vec{\tau}_{i}$.
Then we have
\begin{eqnarray}
(II)&=&
-\sum_{t=nlm}[\frac{1}{N_{cell}\Omega_{cell}}\sum_{ij}e^{-i(\vec{k'}-\vec{k})\cdot(\vec{R}_{j}+\vec{\tau}_{i})}][\int
d\vec{r}_{1}'e^{-i\vec{k'}\cdot\vec{r}_{1}'}\psi_{nlm}(r_{1}',\theta_{1},\phi_{1})] \notag\\
&& \times[\int
d\vec{r}_{2}'\psi_{nlm}^{*}(r_{2}',\theta_{2},\phi_{2})\xi(r_{2}')\vec{l}(\vec{r}_{2}')e^{i\vec{k}\cdot\vec{r}_{2}'}].
\label{ii1}
\end{eqnarray}
We denote the first square bracket in Eq. (\ref{ii1}) by [A], the
second one by [B] and the third square bracket by [C]. Each of these
terms will now be individually analyzed. [A] includes the local
position information of all atoms in a unit cell, as given by
\begin{equation}
[A]=\frac{1}{\Omega_{cell}}\sum_{i}e^{i\vartriangle\vec{k}\cdot\vec{\tau}_{i}},\\
\label{inte1}
\end{equation}
where $\vartriangle\vec{k}=\vec{k'}-\vec{k}$ is the change in the
wave vector. When we consider single-element crystal, there is only
one kind of atom. If we set $n_{cell}$ as the number of all atoms in
a unit cell, then we have
\begin{eqnarray}
[A]&=&\frac{1}{n_{cell}}(\frac{n_{cell}}{\Omega_{cell}})\sum_{i}e^{i\vartriangle\vec{k}\cdot\vec{\tau}_{i}},
\notag \\
&=& \frac{S(\vartriangle\vec{k})}{\Omega_{atom}}, \label{inte12}
\end{eqnarray}
where
\begin{equation}
S(\vartriangle\vec{k})=\frac{1}{n_{cell}}\sum_{i}e^{i\vartriangle\vec{k}\cdot\vec{\tau}_{i}}
\\
\label{sf}
\end{equation}
is the counterpart of the structure factor in Ref. \cite{weisz}, and
$\Omega_{atom}$ is the atomic volume in a unit cell so that
$\Omega_{atom}=\Omega_{cell}/n_{cell}$.

The second square bracket [B] in Eq. (\ref{ii1}) is
\begin{eqnarray}
[B]&=&\sum_{l'=0}^{\infty}C_{l'}\int
d\vec{r}j_{l'}(k'r)Y_{l'k'}^{*}(\theta,\phi)R_{nl}(r)Y_{l}^{m}(\theta,\phi)
\notag \\
&=&\sum_{l'=0}^{\infty}C_{l'}[\int
r^{2}j_{l'}(k'r)R_{nl}(r)dr][\int_{0}^{2\pi}d\phi\int_{0}^{\pi}\sin\theta
Y_{l'k'}^{*}(\theta,\phi)Y_{l}^{m}(\theta,\phi)d\theta] \label{a1}
\end{eqnarray}
where $C_{l'}=(-1)^{l'}i^{l'}[4\pi(2l'+1)]^{1/2}$, $Y_{l}^{m}$ are
spherical harmonics and $j_{l'}$ are spherical Bessel functions. We
use the formula
$e^{i\vec{k}\cdot\vec{r}}=\sum_{l=0}^{\infty}i^{l}[4\pi(2l+1)]^{1/2}j_{l}(kr)Y_{lk}$
and $Y_{lk}(\theta,\phi)$ are the spherical harmonic with the
rotational index $m=0$ in any coordinate system with $z$ in the $k$
direction. Because of the orthogonality of spherical harmonics, the
second square bracket in Eq. (\ref{a1}) gives
$\delta_{ll'}\delta_{mm'}$, i.e., $\delta_{ll'}\delta_{m0}$. Hence
Eq. (\ref{a1}) simplifies to
\begin{equation}
[B]=(-1)^{l}\Omega^{1/2}B_{nl,1}(\vec{k'})\delta_{m,0},\\
\label{a2}
\end{equation}
where
$B_{nl,1}=\Omega^{-1/2}\int_{0}^{\infty}i^{l}[4\pi(2l+1)]^{1/2}j_{l}(k'r)R_{nl}(r)r^{2}dr$
is the counterpart of B function defined in Ref. \cite{weisz},
$R_{nl}$ is the radial part of core wave function.

The third square bracket [C] in Eq. (\ref{ii1}) as given by
\begin{eqnarray}
\Omega^{-1/2}[C]&=& \Omega^{-1/2}\int
d\vec{r}\psi_{nlm}^{*}(r,\theta,\phi)\xi(r)\vec{l}(\vec{r})e^{i\vec{k}\cdot\vec{r}}
\notag \\
&=&\sum_{l'=0}^{\infty}[\Omega^{-1/2}C_{l'}\int
R_{nl}(r)\xi(r)j_{l'}(kr)r^{2}dr]\times[\langle
Y_{l}^{m}|\vec{l}|Y_{l'k}\rangle] \notag \\
&=&
\sum_{l'=0}^{\infty}[radial\hspace{0.3cm}part]\times[angular\hspace{0.3cm}part]
\label{b1}
\end{eqnarray}
where angular part is
\begin{equation}
\langle Y_{l}^{m}|\vec{l}|Y_{l'k}\rangle=\langle
Y_{l}^{m}|\vec{l}|Y_{lk}\rangle\delta_{ll'},\\
\label{ylm}
\end{equation}
Here $m$ in spherical harmonics of angular part must be constrained
by $\delta_{m0}$ in Eq. (\ref{a2}). The same core states
characterized by $nlm$ are expressed in the different coordinates.
The term $\langle\vec{r}_{1}|b_{t}\rangle$ in Eq. (\ref{ii}) means
the core states are projected into $\vec{r}_{1}$ coordinates system
in which $z$ direction is in $\vec{k'}$. Similarly in $\langle
b_{t}|\vec{r}_{2}\rangle$, they are projected into $\vec{r}_{2}$
(i.e. $\vec{r}_{3}$) coordinates, in which $z$ direction is in
$\vec{k}$. So we can write $Y_{l}^{m}$ in Eq. (\ref{ylm}) as
$Y_{lk'}$. Then we get the angular part to be \cite{weisz}
\begin{equation}
\langle
Y_{lk'}|\vec{l}|Y_{lk}\rangle=-i[\frac{dP_{l}(\cos\alpha)}{d(\cos\alpha)}]\frac{\vec{k'}\times\vec{k}}{k'k},\\
\label{ymatrix}
\end{equation}
where $P_{l}$ are the associated Legendre polynomials, $\alpha$ is
the mutual angle between the vector $\vec{k}$ and $\vec{k'}$. From
Eq. (\ref{ymatrix}), (\ref{ylm}) and (\ref{b1}), we get
\begin{equation}
\Omega^{-1/2}[C]=iA_{nl}^{'}[\frac{dP_{l}(\cos\alpha)}{d(\cos\alpha)}]\frac{\vec{k'}\times\vec{k}}{k'k},\\
\label{b2}
\end{equation}
where
$A_{nl}^{'}=\Omega^{-1/2}\int_{0}^{\infty}i^{l}[4\pi(2l+1)]^{1/2}R_{nl}(r)\xi(r)j_{l}(kr)r^{2}dr$,
which is similar to the "A" function in Ref. (\cite{weisz}). Then we
can get the term $(II)$ as
\begin{equation}
(II)=(-i)S(\vartriangle\vec{k})\frac{\vec{k'}\times\vec{k}}{k'k}[\sum_{nl}(-1)^{l}B_{nl,1}(\vec{k'})A_{nl}^{'}(\vec{k})\frac{dP_{l}(\cos(\alpha))}{d\cos(\alpha)}],\\
\label{ii2}
\end{equation}
where the summations shall be taken over the quantum number $n$ and
$l$.

The other terms in (A3) in Ref. \cite{weisz} can also derive from
the similar calculations. Then the final expression of SOC can be
written as
\begin{equation}
H_{\vec{k'}s',\vec{k}s}^{SO}=-\vec{\sigma}_{s's}\cdot(\vec{k'}\times\vec{k})[iS(\vartriangle\vec{k})\lambda_{atom}],\\
\label{hso}
\end{equation}
where $\vec{\sigma}_{s's}=\langle s'|\vec{\sigma}|s\rangle\ $ and
\begin{equation}
\lambda_{atom}=(1/kk')\sum_{nl}(-1)^{l}[B_{nl,1}(\vec{k'})A_{nl}^{'}(\vec{k})+A_{nl}(\vec{k'})B_{nl,1}(\vec{k})
-\sum_{n'}B_{n'l,1}(\vec{k'})B_{nl,1}(\vec{k})\eta_{nn'l}]\frac{dp_{l}(\cos\alpha)}{d(\cos\alpha)},\\
\label{lambatom}
\end{equation}
where the first term comes from the plane wave (p)-core state (c),
the second term comes from c-p, and the third comes from c-c term.
\begin{equation}
\eta_{nn'l}=\int_{0}^{\infty}R_{n'l}(r)\xi(r)R_{nl}(r)r^{2}dr \\
\label{eta}
\end{equation}
is about the spin orbit splitting of isolated atoms. It is just
determined by local wave functions of core states. We shall discuss
two cases.

\textit{Case one: two different atoms in a unit cell} We set
$\gamma=\vartriangle\vec{k}\cdot\vec{\tau}$, $\lambda_{1}$ and
$\lambda_{2}$ are corresponding to the two atoms. We get \cite{che}
\begin{equation}
H_{\vec{k'}s',\vec{k}s}^{SO}=-i\vec{\sigma}_{s's}\cdot(\vec{k'}\times\vec{k})[\lambda^{S}\cos\gamma+i\lambda^{A}\sin\gamma],\\
\label{hso2}
\end{equation}
where $\lambda^{S,A}=(\lambda_{1}\pm\lambda_{2})/2$ are the
symmetric (antisymmetric) contributions to the spin-orbit
Hamiltonian.

\textit{Case two: multi-type and many atoms in a unit cell} If we
have $m$ species atoms in a unit cell, and $n_{1}$, $n_{2}$,
$\cdots$, $n_{m}$ are numbers of atoms for each type atom in a unit
cell. We have $\sum_{i=1}^{m}n_{i}=n_{num}$.
\begin{equation}
H_{\vec{k'}s',\vec{k}s}^{SO}=-i\vec{\sigma}_{s's}\cdot(\vec{k'}\times\vec{k})\sum_{j=1}^{m}S_{j}(\vartriangle\vec{k})\lambda_{j},\\
\label{hso3}
\end{equation}
where $\lambda_{j}$ is for $j$-th species, and structure factor is
$S_{j}(\vartriangle\vec{k})=n_{num}^{-1}\sum_{n_{j}}\exp(i\vartriangle\vec{k}\cdot\vec{\tau}_{n_{j}})$,
here summation over $n_{j}$ means over the atoms of $j$-th species.

If we only keep the third term c-c in Eq. (\ref{lambatom}) (as in
Refs. \cite{che,walter,bloom}) and only consider the main
contribution from the outermost $l=1$ core state, i.e. $p$ orbit.
Also note the relation $dP_{l}/d(\cos\alpha)=1$, for $l=1$, we have
\begin{equation}
\lambda_{atom}=(kk')^{-1}B_{n1,1}(\vec{k'})B_{n1,1}(\vec{k})\eta_{nn1},
\\
\label{lamb1}
\end{equation}
where $n$ is selected as the quantum number characterizing the
outermost $p$ core state, and $\eta_{nn1}$ is spin-orbit parameter
for various atoms, we write it as $\eta_{atom}$. If we redefine the
$B$ function by introducing a normalization constant $C$ as
\begin{equation}
B_{nl}(\vec{k})=C\int_{0}^{\infty}j_{nl}(kr)R_{nl}(r)r^{2}dr, \\
\label{rdb}
\end{equation}
where $C$ is determined by the limit $\lim_{k\rightarrow
0}k^{-1}B_{nl}(k)=1$. And in this procedure, an empirical adjustable
parameter $\mu$ and $\alpha$ describing the ratio of the spin-orbit
contributions for free atoms $A$ and $B$, see Refs. \cite{che}. To
evaluate the $B$ function explicitly, we may determine the
expressions of the core state wave function $R_{nl}(r)$ and we use
the formula in Ref. \cite{clementi}
\begin{equation}
\psi_{nlm}=\sum_{p}C_{nlp}\chi_{plm},\\
\label{psinlm}
\end{equation}
where $C_{nlp}$ are coefficients of expansion, and $\chi_{plm}$ are
Slater-type orbits with integer quantum numbers, namely
\begin{equation}
\chi_{plm}(r,\theta,\phi)=\tilde{R}_{lp}(r)Y_{l}^{m}(\theta,\phi), \\
\label{chi}
\end{equation}
where
\begin{equation}
\tilde{R}_{lp}=[(2n_{lp})!]^{-1/2}(2\zeta_{lp})^{n_{lp}+1/2}r^{n_{lp}-1}e^{-\zeta_{lp}r},\\
\label{rlp}
\end{equation}
where $C_{nlp}$ and $\zeta_{lp}$ can be found in the tables in Ref.
\cite{clementi}. So the functions $\psi$ in Eq. (\ref{psinlm}) are
\begin{eqnarray}
\psi_{nlm}&=&[\sum_{p}\tilde{R}_{lp}(r)C_{nlp}]\times
Y_{l}^{m}(\theta,\phi) \notag\\
&=& R_{lp}(r)Y_{l}^{m}(\theta,\phi), \label{phinlm}
\end{eqnarray}
In terms of these functions, we can calculate the integral as
\begin{eqnarray}
I(k)&=& \int_{0}^{\infty}j_{l}(kr)R_{nl}(r)r^{2}dr \notag \\
&=&
\sum_{p}C_{nlp}\int_{0}^{\infty}j_{l}(kr)\tilde{R}_{lp}(r)r^{2}dr
\notag \\
&=& \sum_{p}C_{nlp}N_{lp}I_{l}^{p}(k), \label{ik}
\end{eqnarray}
where $N_{lp}=[(2n_{lp})!]^{-1/2}(2\zeta_{lp})^{n_{lp}+1/2}$ and
$I_{l}^{p}(k)=\int_{0}^{\infty}j_{l}(kr)r^{n_{lp}+1}e^{-\zeta_{lp}r}dr$.
The integral of $I_{l}^{p}(k)$ can be easily derived when $l=1$. It
is
\begin{equation}
I_{l=1}^{p}(k)=\frac{(n_{lp}-1)!(\sin(n_{lp}\nu)-\frac{kn_{lp}\cos[(n_{lp}+1)\nu]}{(\zeta_{lp}^{2}+k^{2})^{1/2}})}{k^{2}(\zeta_{lp}^{2}+k^{2})^{n_{lp}/2}},\\
\label{il1}
\end{equation}
where $\nu=\arctan(k/\zeta_{lp})$. If we calculate the value of the
limit $k\rightarrow 0$, it shall be pointed out that the spherical
Bessel functions $j_{l}(kr)$ can be formulated
$j_{l}(x)=\sqrt{\frac{\pi}{2x}}J_{l+1/2}(x)$, where $J_{l+1/2}(x)$
is a Bessel function of the first kind. Then we can get
\begin{equation}
\lim_{k\rightarrow
0}I_{l=1}^{p}(k)=\frac{k}{3}\frac{(n_{lp}+2)!}{\zeta_{lp}^{n_{lp}+3}}.\\
\label{ik0}
\end{equation}
This result is used to derive the normalization constant in Eq.
(\ref{rdb}).


\end{document}